# Theoretical studies on anisotropic charge mobility, band structure, and non-linear optical calculations of ambipolar type organic semiconductors

Smruti R. Sahoo[1], Rudranarayan Khatua[1], Suryakanti Debata[1], Sagar Sharma[2], Sridha Sahu[1,*]

[1] High Performance Computing Lab, Department of Physics, Indian Institute of Technology (Indian School of Mines), Dhanbad, Jharkhand-826004, India
[2] Department of Chemistry, School of Fundamental and Applied Sciences, Assam Don Bosco University, Tapesia Gardens, Guwahati, Assam-782402, India

[*]Corresponding author Email: sridharsahu@iitism.ac.in

### Highlights

- Two phenancene's family compounds having identical composition but different molecular packing patterns in crystals.
- Anisotropic charge carrier mobilities have been calculated.
- Molecular packing modes play a significant role in the charge mobility of organic crystals.
- Ambipolar charge transport type of organic semiconductors (OSCs) with better n-type characteristics.
- Calculated electron mobility of **TBT** is very larger (by the order of 1.82 cm$^2$V$^{-1}$s$^{-1}$) than that of **DBP**.
- Possess better air-stability than pentacene molecule and also are chemically stable.
- Band structure and density of states calculations of the studied compounds have been investigated.

### Abstract

The anisotropic charge carrier mobilities of two phenancene series compounds such as dibenzo[a,c]picene (**DBP**) and tribenzo[a,c,k]tetraphene (**TBT**) is investigated based on the first-principle calculations and Marcus-Hush theory. The molecular packing patterns in organic crystal play an important role for determing the charge carrier mobility and hence the device efficiencies designed from the organic materials. Among the studied molecules, **TBT** shows a maximum anisotropic hole ($\mu_h$=0.129 cm$^2$V$^{-1}$s$^{-1}$) and electron ($\mu_h$=1.834 cm$^2$V$^{-1}$s$^{-1}$) mobility, hence possesses an ambipolar semiconducting character. The frontier molecular orbital analyses proved the better air-stability of the studied compounds than the conventional pentacene, because of their higher HOMO energy levels. Band structure calculations of the studied compounds have also been investigated. From non-linear optical (NLO) properties anysis, we found the **TBT** compound shows more NLO response than **DBP**.

### Introduction

In the last two decades, oligoacenes such as tetracene, pentacene and their derivatives have attracted a much attentions because of their potential applications as active layers of organic field-effect transistors (OFETs) [1–3]. The thin-film transistor mobility of pentacene is reported about 3 cm$^2$V$^{-1}$s$^{-1}$, which exceeded the mobility of conventional amorphous silicon (0.5 cm$^2$V$^{-1}$s$^{-}$1) [4,5]. Further, in case of organic semiconductors (OSCs) the highest mobility has been reported as 35 cm$^2$V$^{-1}$s$^{-1}$, for pentacene single crystal [6]. However, the major disadvantage is that acenes are generally unstable in the ambient environment conditions (upon exposure to oxygen and light) [7,8]. It has been also seen that, adding more phenyl rings in the acenes for improving the charge mobility, resulted more degradation and device inefficiency [8]. The high-lying highest occupied molecular orbitals (HOMOs) make them chemically unstable and decrease the solubility and hence limit the large-scale synthesis

[9,10]. Hence, the development of OSCs with higher charge mobility along with better air-stability is still a challenge for the researchers.

Recently, phenancene series such as; phenanthrene, chrysene, picene, fulminene, etc. because of their deeper HOMO energy level as compared to acenes, have emerged as a promising materials with enhanced air-stability, and hence a great point of attractions [11]. For example, picene as a derivative of pentacene has a better stability than the latter and exhibited the maximum single crystal mobility of 2.63 $cm^2V^{-1}s^{-1}$ [12]. Oyama *et al.* studied the opto-electronic and charge transport properties of picene-type π-systems and obtained the high charge carrier mobilities up to 4.7 $cm^2V^{-1}s^{-1}$ [13]. Similarly other reports showed picene as better than amorphous silicon with mobility of 1.40 - 3.2 $cm^2V^{-1}s^{-1}$ [14,15]. Among the other phenancene series, [6] phenancene based thin-film led to mobility of 7.4 $cm^2V^{-1}s^{-1}$, while the thin film single crystal FET designed from [7]phenancene and [8]phenancene were attained a hole mobility of 6.9 and 16.4 $cm^2V^{-1}s^{-1}$, respectively [16,17]. Similarly, Nguyen *et al.* reported the phenancene series such as [6]→[10]phenancene with higher charge mobility and air stability for use in highly efficiency electronic devices [18]. Lamport *et al.* carried out an investigation on the performance of organic field-effect transistors (OFETs) and obtained a maximum FET mobility of 0.3 $cm^2V^{-1}s^{-1}$ for 7,14-bis(trimethylsilylethynyl) benzo[k]tetraphene [19]. In addition, Khatua *et al.* theoretically studied the anisotropic charge mobility of some chrysene derivatives, and reported the maximum hole and electron mobility up to 0.739 and 0.709 $cm^2V^{-1}s^{-1}$, respectively [20].

For organic semiconducting crystals, the charge transport is sensitive to the molecular disorders, packing modes, and the intermolecular π-orbital overlap between the neighbouring molecules [21,22]. The variation in the intermolecular π-orbital interactions can affect the charge transport anisotropies and efficiencies. For example, the effect of crystal packing on the organic thin-film transistor performance of guest-host system based on the syn and anti isomers of triethylsilylethynyl anthradithiophene (TES ADT) is systematically acomplished by Hailey *et al.* [23]. Shi *et al.* theoretically investigated that, the understanding of molecular packing motifs and intermolecular interactions can help into the design novel organic semiconducting materials [24]. Further, a comprehensive overview of the intermolecular packing, morphology, and charge transport features of various organic semiconductors were successfully reported by Wang *et al.* [25]. Some authors also successfully studied the effect of functionalization (e.g. halogens), influence of nitrogen positions on the molecular packing, electronic coupling, and charge mobility of organic semiconductors [26,27].

In this study, we investigate the geometric structures, reorganization energies, molecular orbitals, ionization potentials, electron affinities, and stability of two phenancene series derivative compounds such as dibenzo[a,c]picene (DBP) and tribenzo[a,c,k]tetraphene (TBT) which are compositionally identicals, but differ in the molecular packings of the crystal by using density functional theory (DFT). The effective intermolecular electronic couplings and anisotropic charge carrier mobilities of the studied compounds were calculated within the framework of first-principles quantum chemical calculations and Marcus-Hush theory. The molecular packings and intermolecular interactions play an important role in the anisotropic nature of charge mobilities of the organic crystals; this has been successfully discussed in this study. We provide a details band structure calculations along with density of states (DOS) of the investigated materials. The non-linear optical (NLO) responses of the materials are also discussed. The calculated higher charge mobility (ambipolar nature) and improved stability of the compounds will be helpful for optoelectronic applications such as organic field-effect transistors (OFETs).

## Theoratical details

The charge carrier mobility is one of the most crucial parameters, which measure the performance of organic electronic devices. The anisotropic charge mobilities of the organic crystals were predicted based on the combination of first-principles quantum mechanics calculations and Marcus-Hush theory [28]. At room temperature, the intermolecular charge transfer rate ($K$) by following the Marcus-Hush theory is written as;

$$K = \frac{V^2}{\hbar} \left(\frac{\pi}{\lambda k_B T}\right)^2 \exp\left(-\frac{\lambda}{4 k_B T}\right) \cdots\cdots (1)$$

where $V$ is the electronic coupling or charge transfer integral, and $k_B$ is Boltzmann constant, and $\lambda$ is the reorganization energy.

Though band model and hopping model are two widely used methodologies to explain charge charge transport mechanism in organic semiconductors, the later is used in the cases of localized charge carriers with intermolecular electron coupling being far less than the electron-vibration coupling [29–31].

Based on the molecular molecular orbitals of the conjugated organic materials, the intermolecular effective electronic coupling, $V_{eff}$ for hole (h) or electron (e) ($V_{eff}^{h/e}$) can be evaluated by using the direct coupling (DC) method as [28,32–34];

$$V_{eff}^{h/e} = \frac{J_{\alpha\beta} - \frac{1}{2} S_{\alpha\beta}(t_{\alpha\alpha}^{H/L} + t_{\beta\beta}^{H/L})}{1 - S_{\alpha\beta}^2} \cdots\cdots (2)$$

where $J_{\alpha\beta}$ and $S_{\alpha\beta}$ are charge transfer integrals and spatial overlaps, respectively. $t_{\alpha\alpha}^{H/L}$ and $t_{\beta\beta}^{H/L}$ are the site energies contributed from highest occupied molecular orbitals (HOMO) and lowest unoccupied molecular orbitals (LUMO) respectively. In direct coupling, the electron dimer state are defined in terms of localized monomer orbitals and the charge-localized monomer monomer diabatic states [35].

Assuming that $H_{KS}$ is the Kohn-Sham Hamiltonian of the dimer system with $t_{\alpha}^{H/L}$ and $t_{\beta}^{H/L}$ being HOMO or LUMO of two constituting monomers α and β, the specified terms can be evaluated as [32,33];

$$J_{\alpha\beta} = \langle \phi_{\alpha}^{H/L} | H | \phi_{\beta}^{H/L} \rangle \cdots\cdots (3)$$

$$S_{\alpha\beta} = \langle \phi_\alpha^{H/L} | \phi_\beta^{H/L} \rangle \cdots\cdots (4)$$

$$t_{\alpha\alpha} = \langle \phi_\alpha^{H/L} | H_{KS} | \phi_\alpha^{H/L} \rangle \cdots\cdots (5)$$

$$t_{\beta\beta} = \langle \phi_\beta^{H/L} | H_{KS} | \phi_\beta^{H/L} \rangle \cdots\cdots (6)$$

At room temperature, assuming that the charge motion is a homogeneous random walk and hopping events are independent of each other, the charge transfer between the adjacent molecules of organic crystal exhibit a diffusive behaviour. According to the Einstein-Smonluchowski relation, the drift mobility of the organic crystal is given by [28,32,33];

$$\mu = \frac{e}{k_B T} D \cdots\cdots (7)$$

where $D$ is the isotropic charge diffusion coefficient and is defined as follows;

$$D = \frac{1}{2n} \sum_i r_i^2 . K_i . P_i \cdots\cdots (8)$$

$n$ represents the spatial dimensionality, $r_i$ defines the intermolecular distance for $i$th hopping pathway, $K_i$ is the charge hopping rate and $P_i$ represents the hopping probability which is calculated as;

$$P_i = \frac{K_i}{\sum K_i} \cdots\cdots (9)$$

For the organic crystals, the value of anisotropic charge mobility are calculated in a certain directions, which depends on the orientation of the crystals. Hence, we analyze the mobility of the studied organic crystals in a specific surface for each directions in terms of angles ($\gamma_i$) between the charge hopping pathways and plane of interest ($K_i.r_i.\cos\gamma_i$). If $\Phi$ is the angle of orientation of the transport channel relative to the reference axis (a, b, or c) and the $\theta_i$ are the angles between the projected hopping pathways, then the mobility orientation function to predict the angular-anisotropic charge carrier mobility in the organic crystals can be deduced from the relation [28];

$$\mu_\Phi = \frac{e}{2k_B T} \sum_i K_i . r_i^2 . P_i . \cos^2 \gamma_i \cos^2(\theta_i - \Phi) \cdots\cdots (10)$$

### Computational details

The initial geometries of the molecules are collected from the reported crystal structures of **DBP** (CCDC 1521686) and **TBT** (CCDC 1521684) listed in the Cambridge Crystallographic Data Centre [36]. The geometries of the molecules are fully optimized in their ground and ionized states by using B3LYP hybrid exchange-correlation functional and 6-31+G(d) basis set. The charge transfer properties such as; frontier molecular orbital (FMO), reorganization energy (λ), ionization potential (IP), and electron affinity (EA) of the studied compounds are calculated by using the same optimization level of theory.

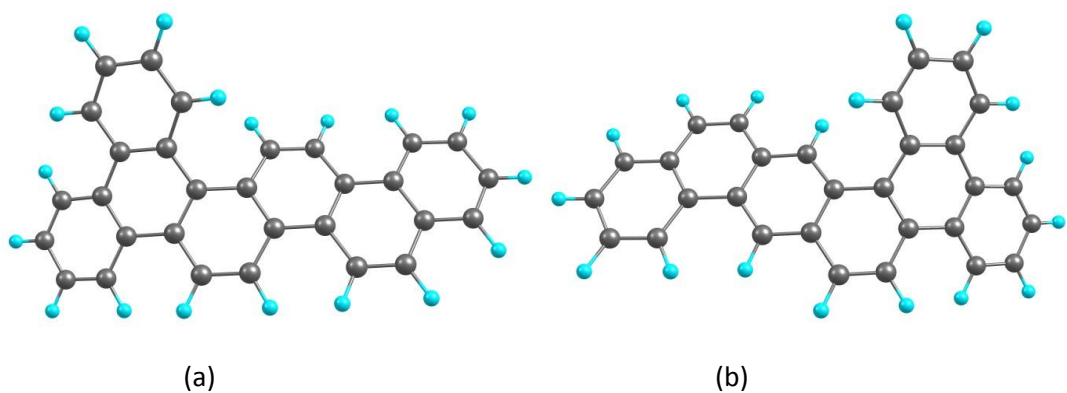

Figure 1: The ground state optimized structures of the molecules (a) Dibenzo[a,c] picene (DBP) and (b) Tribenzo[a,c,k] tetraphene (TBT).

Further, the calculations of electronic coupling parameters such as; effective transfer integral ($V_{eff}$), site energy (t), and spatial overlap (S) were executed with the help of AOMix program by using Kohn-Sham Hamiltonian ($H_{KS}$) [28,32,33,37] through fragment molecular orbital approach of dimers by using the local density functional with PW91 generalized gradient approximations, and 6-31G* basis set [37]. All the quantum chemical calculations were performed by using the computational chemistry package, Gaussian 09 [38]. Further, the density of state (DOS) and electronic band-structure of the studied crystals DBP and TBT were calculated by using the first-principle calculation based on DFT and PBE-GGA level of theory [39], and the simulations were performed with the help of Vienna package Wien2K [40]. We use the Monkhorst-Pack method to generate k-point meshes 1x3x1 and 3x1x1 for monoclinic and orthorhombic unit cell, respectively [41]. In addition, the non-linear optical properties of the studied crystals are calculated at B3LYP and CAM-B3LYP with 6-31+G* basis set levels of theory.

## Result and discussion

### Structure

The investigated two conjugated compounds consist of seven fused benzene rings and the compounds are named as; dibenzo[a,c]picene (**DBP**) and tribenzo[a,c,k]tetraphene (**TBT**) respectively. Though these compounds are compositionally identical ($C_{30}H_{18}$), however they differ in their crystallographic structures, parameters, and packing patterns. The **DBP** compound has the crystallographic cell parameters such as a=7.560 Å, b=20.153 Å, c=23.743 Å, α=β=γ=90° and space group Pbca (orthorhombic); whereas **TBT** possesses the cell parameters as a=12.778 Å, b=5.158 Å, c=14.363 Å, α=γ=90°, β=98.059° and space group P21 (monoclinic) [36]. The optimized ground state structure of **DBP** and **TBT** are displayed in Figure 1. Similar to the reported crystal structures, the optimized structures show some twisting in the molecular backbone which is because of the steric interaction between the benzo moiety and the fulminene ([6]phenacene) framework [36]. Further, the optimized ground state geometries of the studied compounds are found to be nearly identical to the reported crystal structure coordinates.

## Frontier molecular orbitals (FMO) and stability

Frontier molecular orbitals consist of highest occupied molecular orbital (HOMO) and lowest unoccupied molecular orbital (LUMO) and these are having great importance for analysing the optical, charge transport and stability of the conjugated organic compounds. The computed electron density distribution in the form of molecular orbitals along with the HOMO/LUMO energy levels and energy gaps of the studied **DBP** and **TBT** compounds are shown in Figure 2. All the molecular orbitals show π- in character.

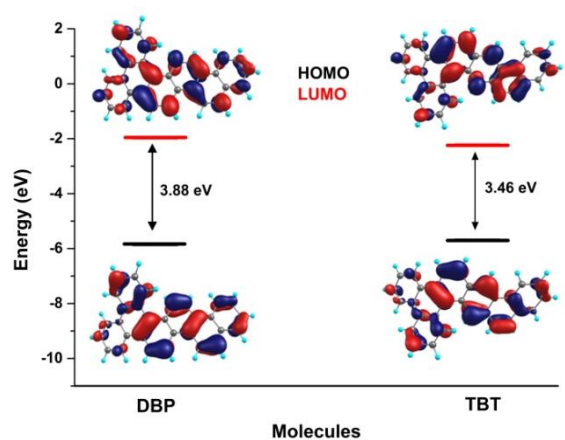

Figure 2: Calculated electron density distribution, HOMO/LUMO energy levels and the energy gap between them of the studied compounds.

The calculated values of the $E_{HOMO}$/$E_{LUMO}$ energy levels, energy gap of **DBP** and **TBT** are -5.65/-1.77 eV, 3.88 eV and -5.46/-2.01 eV, 3.46 eV respectively. The energy gap is reduced by 0.42 eV in **TBT** as compared to **DBP** molecule. From Figure 2 it is seen that, the electron density distributions of both HOMO and LUMO are delocalized over the whole molecules with more delocalization at the centre. The calculated HOMO energies of the molecules are found to the desired -5.2 eV below the vacuum level, and further these relatively deep HOMO levels and large band gaps infer the good ambient air-stability of these compounds [9,42]. In addition, the decrease of HOMO-LUMO energy gap in **TBT** indicates that, **TBT** owns excellent electron transport performance; whereas the higher $E_{HOMO}$ energy level also induces the better hole injection ability in **TBT** molecule.

## Reorganization energies

Reorganization energies is treated as an important parameter which influences the charge mobility of the organic semiconductors. The reorganization energies of the studied compounds were calculated from the potential energy surfaces in both adiabatic and vertical form at B3LYP/6-31+G(d) theory of level and the values are provided in Table 1. Here, we considered that the contribution of the external environment to the reorganization energy is weak and is thereby neglected. The values of the internal reorganization energies $\lambda_h$/$\lambda_e$ of the investigated compounds **DBP** and **TBT** are 0.147/0.186 eV and 0.138/0.172 eV, respectively. It is reported that the larger reorganization energy values is not good for higher charge transport [43]. From the calculations, it is observed that both hole ($\lambda_h$) and electron reorganization energy ($\lambda_e$) of both **DBP** and **TBT** compounds are nearly same, however in case of **TBT**, a small reduced energy values (for example, $\lambda_h$ is reduced by 0.009 eV and $\lambda_e$ by 0.014 eV) is observed as compared to **DBP** compound.

Table 1: Calculated reorganization energies, ionization potentials, electron affinities, HOMO/LUMO energy levels and the energy gaps of the studied compounds.

| Molecules | $\lambda_h$ (eV) | $\lambda_e$ (eV) | $IP_a$ (eV) | $IP_v$ (eV) | $EA_a$ (eV) | $EA_v$ (eV) | $E_{HOMO}$ (eV) | $E_{LUMO}$ (eV) | Energy gap (eV) |
|---|---|---|---|---|---|---|---|---|---|
| DBP | 0.147 | 0.186 | 6.77 | 6.84 | 0.69 | 0.60 | -5.65 | -1.77 | 3.88 |
| TBT | 0.138 | 0.172 | 6.60 | 6.67 | 0.92 | 0.83 | -5.47 | -2.01 | 3.46 |

### Ionization potential and electron affinity

Ionization potential (IP) and electron affinity (EA) are the important parameters for optoelectronic applications of conjugated organic materials. It is reported that, a smaller IP value facilitates the hole injection from HOMO, whereas a larger EA vlaue will ease the electron injection into the empty LUMO of the semiconducting materials. The IP and EA of the investigated compounds are calculated from the potential energy surfaces at B3LYP/6-31+G(d) level and the values are summerized in Table 1. The calculated $IP_a/IP_v$ values for **DBP** and **TBT** are 6.77/6.84 eV and 6.60/6.67 eV; whereas the $EA_a/EA_v$ values are 0.69/0.60 eV and 0.92/0.83 eV, respectively. Hence, we observed a smaller IP values (as it is reduced by a factor of 0.17 eV) in case of **TBT** as that from **DBP** compound, which will enhance hole injection and incease the p-type characteristics in the former. Similarly, the compound **TBT** exhibits larger $EA_a/EA_v$ values (increased by the order of 0.23 eV) as that of **DBP**, which may be good for the electron transport and behaves as n-type semiconducting materials.

It is also reported that, the ionization potential and electron affinity are closely related to the HOMO and LUMO energy levels of the materials such as; IP=-$E_{HOMO}$ and EA=-$E_{LUMO}$, respectively. In addition, the stability is a useful standard to measure the nature if devices for optoelectronic applications, and which in further dierctly proportional to the hardness ($\eta$) of the materials. The $\eta$ factor can be calculated as $\eta$=(IP-EA)/2 and the calculated values are 1.94 eV and 1.73 eV for **DBP** and **TBT** compounds respectively. Hence, the stability of the studied compounds follows the order **DBP** > **TBT**.

### Electronic coupling

The molecular packing mode is of great importance for the electronic coupling and charge transport in conjugated organic crystals [33,34,44]. It is seen that, electronic coupling between the same organic molecular stacking layer is much stronger than that between the molecules in two adjacent stacking molecular layers. Hence, in this study we have considered only the same stacking molecular stacking layer and a particular crystallographic plane for the calculation of electronic coupling (V) of organic materials. The molecular packing modes along with all nearest possible projected hopping pathways of the studied crystals are displayed in Figure 3. It is clear from the Figure 3 that the packing structure of both **DBP** and **TBT** show herringbone patterns and we noticed only two types of dimers in the crystal such as, parallel (P) or face-to-face and transverse (T) or face-to-edge dimers. The values of hopping distances between the dimers and the angle between the hopping pathways and reference axis of the crystal of both the crystals are depicted in Figure 3. The intermolecular distances corresponding to the hopping pathways $P_1$, $T_1$, $T_2$, $T_3$, $T_4$, and $T_5$ are denoted as $r_{P1}$, $r_{T1}$, $r_{T2}$, $r_{T3}$, $r_{T4}$, and $r_{T5}$ and the corresponding hopping angles are $\theta_{P1}$, $\theta_{T1}$, $\theta_{T2}$, $\theta_{T3}$, $\theta_{T4}$, and $\theta_{T5}$ respectively.

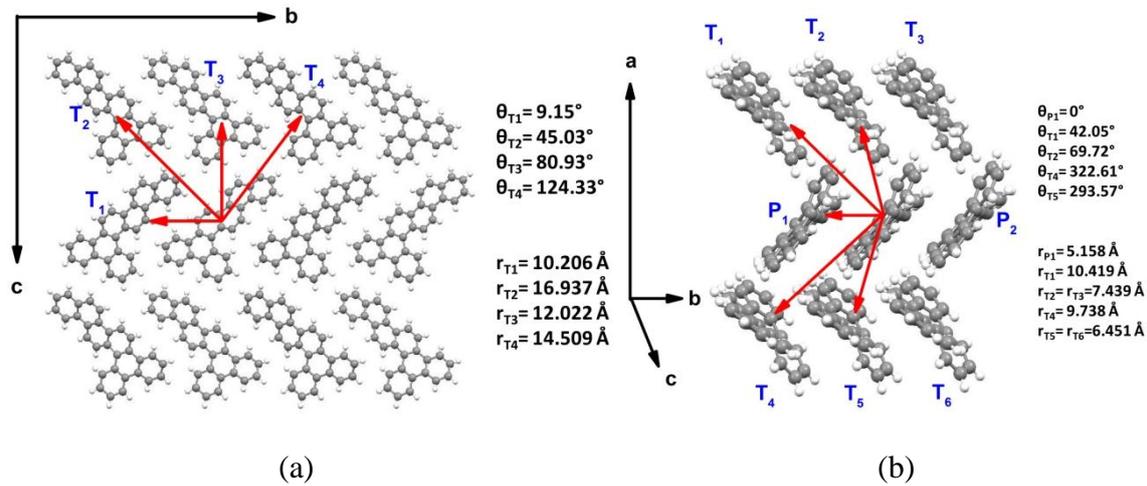

(a)          (b)

**Figure 3:** Schematic daigram of molecular packing modes, crystallographic plane, projected charge hopping pathways, intermolecular center-of-mass distances, and angle of projected hopping pathways relative to reference axis of (a) DBP and (b) TBT.

(Note: For TBT compound, the dimers like $T_2$, $T_3$ and $T_5$, $T_6$ are similar to each other. Hence, we have considered only T2 and T5 dimer.)

The calculated values of spatial overlap (S), site energy (t), and effective transfer integral ($V_{eff}$) of both **DBP** and **TBT** crystals at PW91/6-31G* level of theory are summarized in the Table 2. For organic semiconducting materials, the electronic coupling (V) parameter plays an important role to assess the charge transport performance. It is known that, smaller intermolecular center-to-center distance and face-to-face parallel packings are the two crucial factors for larger electronic coupling. For example, in case of **DBP** compound the $T_1$ pathway with hopping distance 10.206 Å resulted the largest effective electronic coupling $V_{eff}^h$ (for hole) and $V_{eff}^e$ (for electron) values such as -5.70 meV and -3.50 meV, respectively. Similarly, as compared to other pathways, the $P_1$ pathway of **TBT** compound led to the largest $V_{eff}^h$ and $V_{eff}^e$ values such as 17.300 meV and -68.307 meV, because of the minimum intermolecular hopping distance (5.158 Å) and face-to-face parallel overlapping (spatial overlap were found maximum as -0.0035 for hole and 0.0103 for electron). The larger values of both hole and electron coupling in **TBT** as compared to **DBP** compound, may lead to better hole and electron mobilities and hence enhance the p-type and n-type characteristics in the former.

### Anisotropic charge mobility

The anisotropic charge mobility of the studied **DBP** and **TBT** compounds are calculated in a particular transistor channel which depends on the specific surface of the crystals. Considering the reorganization energies and the effective intermolecular electronic coupling, the anisotropic charge mobilities were computed by using the Equation... , and shown in Figure 4 and the mobility ranges are listed in the Table 2. It can be noticed that the simulated angular hole and electron mobility show notable anisotropic nature. As it is already been discussed, the maximum angular hole and electron mobilities were noticed in directions of smaller hopping distances and for face-to-face parallel packing pathways due to large hole and electron intermolecular coupling.

Table 2: Calculated effective transfer integrals ($V_{eff}$), spatial overlap (S), site energies (t), and the range of anisotropic hole and electron mobility of the studied compounds.

| Compounds | Dimers | $V_{eff}^h$/ $V_{eff}^e$ (meV) | $S_{\alpha\beta}^h$/ $S_{\alpha\beta}^e$ | $t_{\alpha\alpha}^H$/ $t_{\alpha\alpha}^L$ (eV) | $t_{\beta\beta}^H$/ $t_{\beta\beta}^L$ (eV) | $\mu_\Phi^h$/ $\mu_\Phi^e$ (cm$^2$V$^{-1}$s$^{-1}$) |
|---|---|---|---|---|---|---|
| **DBP** | T$_1$ | -5.70/-3.50 | 0.0005/0.0003 | -4.807/-2.024 | -4.875/-2.092 | (0.0491 - 0.0236)/ (0.0154 - 3.68x10$^{-6}$) |
| | T$_2$ | 0.00/0.00 | 0.0/0.0 | -4.826/-2.043 | -4.811/-2.028 | |
| | T$_3$ | -4.80/0.40 | 0.0005/0.0 | -4.811/-2.028 | -4.864/-2.080 | |
| | T$_4$ | 0.20/-0.10 | 0.0/-0.0001 | -4.834/-2.051 | -4.836/-2.053 | |
| **TBT** | P$_1$ | 17.300/-68.307 | -0.0035/0.0103 | -4.600/-2.196 | -4.581/-2.173 | (0.1294 - 0.0389)/ (1.834 - 1.19x10$^{-6}$) |
| | T$_1$ | 0.00/0.00 | 0.0/0.0 | -4.598/-2.191 | -4.678/-2.270 | |
| | T$_2$ | -0.90/-0.20 | 0.0001/-0.0001 | -4.524/-2.118 | -4.752/-2.339 | |
| | T$_4$ | 0.00/0.00 | 0.0/0.0 | -4.687/-2.278 | -4.590/-2.182 | |
| | T$_5$ | -12.50/0.70 | 0.0014/0.0004 | -4.793/-2.389 | -4.452/-2.042 | |

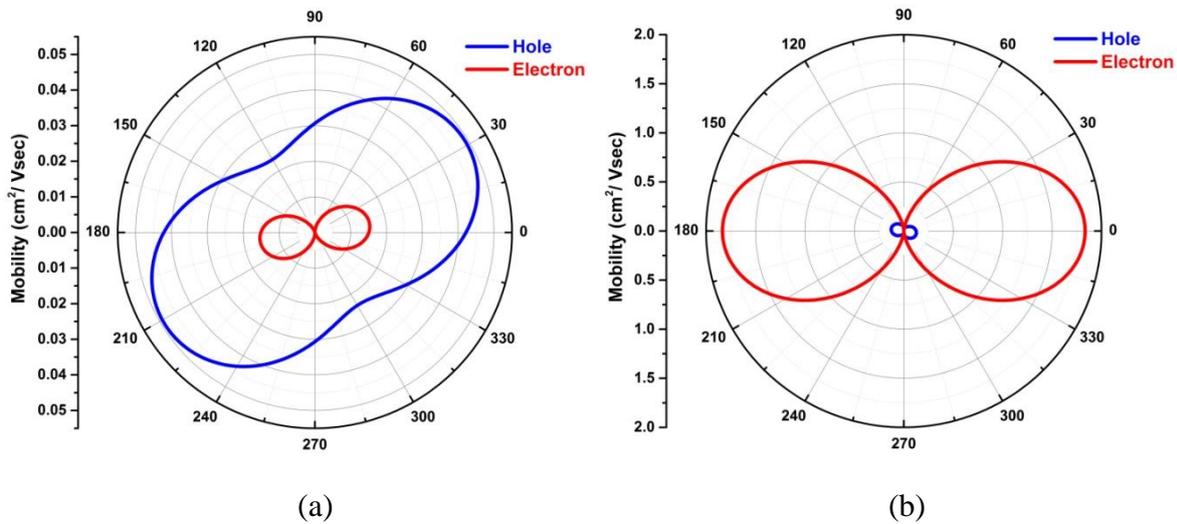

(a) (b)

Figure 4: Simulated angular anisotropic hole and electron mobility of the studied compounds (a) DBP and (b) TBT.

The compound **TBT** presents the largest charge mobility values as compared to **DBP**, and the maximum electron mobility ($\mu_\Phi^e$) is 1.834 cm$^2$V$^{-1}$s$^{-1}$ at $\Phi$=0°/180° and the maximum hole mobility ($\mu_\Phi^h$) is 0.129 cm$^2$V$^{-1}$s$^{-1}$ at $\Phi$=167.87°/347.78°. These maximum values correspond to the face-to-face parallel pathway P. The larger electron mobility assures the better electron transport (n-type) performance of the material. Similarly, the maximum $\mu_\Phi^h$ and $\mu_\Phi^e$ values for **DBP** was found to be 0.049 cm$^2$V$^{-1}$s$^{-1}$ at $\Phi$=31.51°/211.42° and 0.0154 cm$^2$V$^{-1}$s$^{-1}$ at $\Phi$=9.16°/189.07°, respectively. This maximum angular electron mobility of **DBP** is consistent with the T$_1$ pathway which having larger electronic coupling value. For compound **DBP**, the minimum angular hole and electron anisotropic mobility were computed as 0.0236 cm$^2$V$^{-1}$s$^{-1}$ at $\Phi$=121.47°/301.37° and 3.68x10$^{-6}$ at $\Phi$=99.12°, respectively. Similarly, the minimum $\mu_\Phi^h$ and $\mu_\Phi^e$ for **TBT** were calculated as 0.0389 cm$^2$V$^{-1}$s$^{-1}$ at $\Phi$=77.92°/257.83° and

1.19x10$^{-6}$ at $\Phi$=89.95°, respectively. Since, the reorganization energies of both these compounds are nearly same, hence the larger effective electronic coupling values and smaller intermolecular hopping distances resulted largest mobilities in case of compound **TBT**.

The calculated anisotropic mobilities demonstrate that, though the studied organic crystals have the same chemical formula and structure; however the molecular packing modes play an important role for measuring the charge transport performance of the materials. Both hole and electron mobilities of **TBT** compound is larger than that of **DBP**, inferring the former to be good candidate as an ambipolar organic semiconducting materials with more n-type characteristics.

## Band structure and density of states

The calculated band structure and total density of states (DOS) of studied **DBP** and **TBT** crystals are depicted in Figure 5. Both band dispersion and band splitting for **DBP** and **TBT** crystals are calculated near the Fermi level. In **TBT** crystal, the conduction band minima (CBM) and valence band maxima (VBM) are found to occur at M-point, which contributes to direct band gap (2.23 eV) semiconductor. The largest band dispersion of CB and VB for TBT are calculated to be 295.7 meV and 118.85 meV along ΓM and ΓR. Similarly, the largest band splitting for the same are 78.12 meV and 47.0 meV at X-point and Γ-point, respectively.

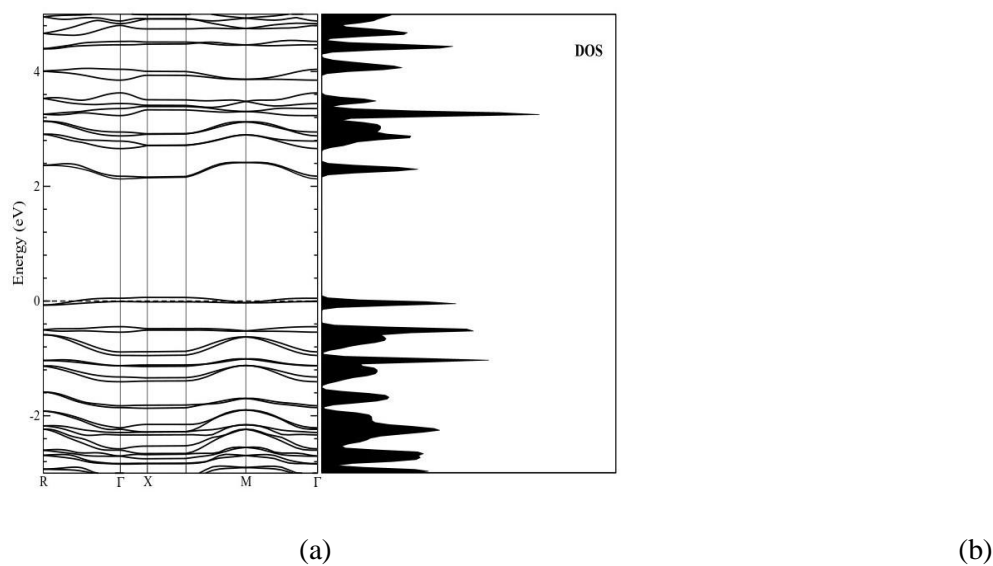

(a)  (b)

Figure 5: Calculated band structure and density of states of the crystals (a) TBT and (b) (DBP).

## Non-linear optical (NLO) properties

The non-linear optical (NLO) parameters such as; the static averaged polarizability ($<\alpha>$), polarizability anisotropy ($\Delta\alpha$), and the first-order hyperpolarizability ($\beta_{tot}$) of studied crystals **DBP** and **TBT** are calculated and summerized in Table 3. It is reported that, the smaller 6-31+G* basis set renders an adequte alternative to the Sadlej's POL basis set for predicting the non-linear optical responses of π-conjugated materials [45,46].

Table 3 : Computed static averaged polarizability, polarizability anisotropy, and first order hyper polarizability of DBT and TBT compounds.

| Compounds | B3LYP/6-31+G(d) | | | CAM-B3LYP/6-31+G(d) | | |
|---|---|---|---|---|---|---|
| | $<\alpha>$ (Å$^3$) | $\Delta\alpha$ (Å$^3$) | $\beta_{tot}$ x 10$^{-53}$ (C$^3$m$^3$J$^{-2}$) | $<\alpha>$ (Å$^3$) | $\Delta\alpha$ (Å$^3$) | $\beta_{tot}$ x 10$^{-53}$ (C$^3$m$^3$J$^{-2}$) |
| **DBP** | 60.56 | 61.83 | 165.5281 | | | |
| **TBT** | 61.64 | 63.81 | 507.5111 | | | |

Hence, we computed the NLO properties of investigated compounds at B3LYP/6-31G+(d) and CAM-B3LYP levels of theory. The $<\alpha>/\Delta\alpha$ of studied compounds **DBP** and **TBT** are calculated as 60.56/61.83 Å and 61.64/63.81Å respectively. It can be seen that, both $<\alpha>$ and $\Delta\alpha$ of **TBT** were enhanced by 1.08 Å and 1.98 Å as compared to **DBP** compound. However, we noticed interesting results in case of first-order hyperpolarizability ($\beta_{tot}$) of investigated compounds. It can be observed that, **TBT** compound yields quite large values of $\beta_{tot}$ (507.5111x10$^{-53}$ C$^3$m$^3$J$^{-2}$) which is about 3 times larger than that of **DBP** (165.5281x10$^{-53}$ C$^3$m$^3$J$^{-2}$). The large value is consistent with tensor analysis of hyperpolarizability, which shows that the values of $\beta_{xxy}$ and $\beta_{xyy}$ components of **TBT** are large than that of **DBP** compound. This further indicates the high displacement of the charge cloud in that particular direction. Therefore, from the analysis we encountered that the **TBT** compound shows more NLO response than that **DBP**.

## Conclusions

The anisotropic charge carrier mobilities of two phenancene series compounds such as dibenzo[a,c]picene (**DBP**) and tribenzo[a,c,k]tetraphene (**TBT**) is investigated based on the first-principle calculations and Marcus-Hush theory. The molecular packing patterns in organic crystal play an important role for determing the charge carrier mobility and hence the device efficiencies designed from the organic materials. Among the studied molecules, TBT shows a maximum anisotropic hole ($\mu_h$=0.129 cm$^2$V$^{-1}$s$^{-1}$) and electron ($\mu_h$=1.834 cm$^2$V$^{-1}$s$^{-1}$) mobility, hence possesses an ambipolar semiconducting character. The frontier molecular orbital analyses proved the better air-stability of the studied compounds than the conventional pentacene, because of their higher HOMO energy levels. Band structure calculations of the studied compounds have also been investigated. From non-linear optical (NLO) properties anysis, we found the **TBT** compound shows more NLO response than **DBP**.